\documentclass[11pt, oneside]{article}   	

\usepackage{tikz}

\pdfoutput=1
\usepackage{jheppub}
\usepackage{comment}

\usepackage[UKenglish]{babel}
\usepackage[T1]{fontenc}
\usepackage{lmodern}

\usepackage{graphicx}	

\usepackage{amsmath}	
\usepackage{physics}

\usepackage{color}
\usepackage{mathtools}
\usepackage{amssymb}
\usepackage{todonotes}

\usepackage{hyperref}

\hypersetup{
    bookmarks=true,%
    colorlinks,%
    citecolor=blue,%
    filecolor=black,%
    linkcolor=blue,%
    urlcolor=blue
}

\usepackage{overpic}
\usepackage{array}
\usepackage{rotating}

\usepackage{amssymb}

\title{Carroll Strings with an Extended Symmetry Algebra}

                                           \author{Matthias Harksen$^{1}$,}
                                            \author{Diego Hidalgo$^{1}$,}
                                           \author{Watse Sybesma$^{1,2,3}$,}
                                           \author{and L\'{a}rus Thorlacius$^{1}$}
                                           \affiliation{$^{1}$Science Institute
                                           University of Iceland \\Dunhaga 3, 107 Reykjav\'{i}k, Iceland}
                                           
                                                                                  \affiliation{$^{2}$Department of Applied Mathematics and Theoretical Physics\\ University of Cambridge, Cambridge CB3 0WA, United Kingdom}

                                            \affiliation{$^{3}$Isaac Newton Institute for Mathematical Sciences\\
                                            University of Cambridge, Cambridge CB3 0EH, United Kingdom}

                                           \emailAdd{mbh6@hi.is}
                                           \emailAdd{dhidalgo@hi.is}
                                           \emailAdd{zhws2@cam.ac.uk}
                                           \emailAdd{lth@hi.is}

\abstract{
Starting from the Polyakov action we consider two distinct Carroll limits in target space, keeping the string worldsheet relativistic. The resulting magnetic and chiral Carroll string models exhibit different symmetries and dynamics. Both models have an infinite dimensional symmetry algebra with Carroll symmetry included in a finite dimensional subalgebra. For the magnetic model, this is the so-called string Carroll algebra. The chiral model realises an extended version of the string Carroll algebra. The magnetic model does not have any transverse string excitations. The chiral model is less restrictive and includes arbitrary left-moving modes that carry transverse momentum but do not contribute to the energy in target space.
}

\begin{document}
\maketitle
\newpage
\section{Introduction}
String theory provides a robust framework for investigating quantum gravity in various decoupling limits. 
In this regard, non-relativistic limits of string theories \cite{Gomis:2000bd,Danielsson:2000gi,Danielsson:2000mu} form a self-contained sector within string theory governed by a relativistic worldsheet metric with Galilean global symmetries on the target spacetime. 
We refer the interested readers to \cite{Oling:2022fft} for a recent review of non-relativistic string theory.
In this note, we consider bosonic string theory in another non-Lorentzian target space limit called the Carroll limit \cite{Levy1965, Bacry:1968zf,Leblonde:1965,Gupta1966OnAA}. 
A pictorial way to highlight the contrasting extremes of these limits is that rather than opening up the 
light cone, as happens in the Galilean limit, the lightcone collapses to a line in the Carroll limit.  
As a result, timelike particles are confined to the temporal axis.
Tachyons still move in the Carroll limit but have no rest frame. 
Furthermore, Carroll boost symmetry requires energy flux to vanish. 
Particles that move are thus forced to have vanishing energy while particles at rest can have non-vanishing energy \cite{deBoer:2017ing,deBoer:2021jej}.
Particle properties in the Carroll limit were further studied in \cite{Bergshoeff:2014jla,Casalbuoni:2023bbh,deboer:2023fnj}, where it was demonstrated that non-trivial dynamics can arise for coupled Carroll particles.
Our primary motivation for developing a Carroll limit of string theory is to see if Carroll string dynamics is less constraining than Carroll particle dynamics. Carroll strings with worldsheet Carroll symmetry were studied in \cite{Isberg:1993av,Bagchi:2015nca,Bidussi:2023rfs,Gomis:2023eav}, whereas target space Carroll symmetry was studied in 
\cite{Cardona:2016ytk,Bagchi:2023cfp,Bidussi:2023rfs}. 
As we proceed, we will comment on similarities and differences between our work and these earlier papers.

We study two inequivalent Carroll limits in the target space of bosonic string theory by introducing different sets of auxiliary worldsheet fields into the Polyakov action. One of these limits corresponds to what is known throughout the literature as a magnetic limit \cite{Duval:2014uoa,Henneaux:2021yzg,deBoer:2021jej,Bergshoeff:2022qkx}. 
Our other Carroll limit breaks worldsheet chirality via a novel auxiliary field configuration. In parallel with the non-relativistic string \cite{Gomis:2000bd}, the Carroll symmetry appears as a global symmetry of the target spacetime, while the worldsheet  remains relativistic. 
The magnetic limit reproduces an action that is invariant under the string Carroll algebra previously derived by Cardona, Gomis, and Pons (CGP) by considering a Carroll limit of the canonical form of the Nambu-Goto action \cite{Cardona:2016ytk}.

The worldsheet dynamics of our magnetic Carroll string model is similar to the CGP model in that no motion is allowed in the transverse directions. The transverse target space momentum is unconstrained, but does not enter into the magnetic Carroll string dispersion relation. This is reminiscent of the behaviour of Carroll particles \cite{Bergshoeff:2014jla,Casalbuoni:2023bbh, deBoer:2021jej}.
The chiral Carroll model also has unconstrained transverse momentum and a simple dispersion relation, where energy does not depend on transverse momentum. Unlike the magnetic model, transverse motion is allowed in the chiral model via arbitrary left-moving string excitations, thus departing from the Carroll particle picture. 
The symmetry algebra of the chiral model is an extension of the string Carroll algebra of the magnetic model. 
Furthermore, an infinite dimensional symmetry extension is realised in both our models in a similar manner as for strings with  Galilean symmetry \cite{Bergshoeff:2019pij,Batlle:2016iel}. 

The remainder of the paper is organised as follows. In Section \ref{sec:mag-carroll}, we establish notation and construct the magnetic Carroll limit by introducing suitable auxiliary fields in the Polyakov action. The symmetry algebra of the magnetic Carroll model and its string dispersion relation are derived. In Section~\ref{sec:chiralcarroll} we present the chiral Carroll model and its extended Carroll string algebra. We then show how to extend the symmetry to an infinite dimensional algebra for both the chiral and magnetic models. 
In Section~\ref{sec:discussions} we summarise our results and discuss some future directions.

\section{Magnetic Carroll string model} \label{sec:mag-carroll}
This section introduces the first of two Carroll limits for strings we consider. The Carroll structure of the model is observed on the target space, while the worldsheet metric remains relativistic. As we will see in Section~\ref{Section:magneticmodelbasics}, the magnetic Carroll string arises from relativistic string theory via an asymmetric scaling of the longitudinal and transverse target space coordinate fields in the Polyakov action that renders the transverse sector `dominant' over the longitudinal sector. The naive divergence that appears when taking this Carroll limit can be controlled by introducing auxiliary fields along the lines of the corresponding construction for non-relativistic strings in \cite{Gomis:2000bd}. The presence of the auxiliary fields severely constrains the dynamics of the magnetic Carroll string model. 

\subsection{Relativistic origin of the magnetic Carroll model}\label{Section:magneticmodelbasics}
Our starting point is the Polyakov action, given by
\begin{equation}\label{eq:polyakov}
	S_{\text{P}}=-\frac{1}{4\pi\alpha'}\int d^{2}\sigma\sqrt{-\gamma}\gamma^{ab}\partial_{a}\hat{X}^{\mu}\partial_{b}\hat{X}^{\nu}\eta_{\mu\nu}
	\,.
\end{equation}
Here lowercase Latin indices run over the worldsheet coordinates $a,b \in \left\{ \sigma, \tau \right\}$ and $\gamma^{ab}$ denotes the inverse of the worldsheet metric.  
The flat $D$-dimensional target-space metric is denoted by $\eta_{\mu \nu}$ with Lorentzian signature. The coupling constant $\alpha^\prime$ is related to the string tension $T$ via $T = 1/(2\pi \alpha^\prime)$.
The worldsheet fields in \eqref{eq:polyakov} are labelled by hats in anticipation of a rescaling to be carried out 
immediately below. We reserve a hat-free notation for the rescaled fields that we work with in the Carroll limit.

We want to take a Carroll limit on the target space while allowing the worldsheet theory to retain its relativistic and Weyl symmetries. In order to achieve this, we introduce an effective coupling constant $\alpha^{\prime}_{\text{eff}}$ and perform an asymmetric scaling of the worldsheet fields by the dimensionless factor of $\alpha^\prime/\alpha^{\prime}_{\text{eff}}$, where longitudinal coordinates in target space, denoted by uppercase Latin indices $A, B = 0,1$, are scaled differently from the transverse coordinates, denoted by uppercase primed indices $A', B' = 2, \ldots, D-1$,
\begin{align} \label{eq:scaling}
    X^{A} = \sqrt{\frac{\alpha'_{\text{eff}}}{\alpha'}}\, \hat{X}^A\,, \hspace{1cm} 
   X^{A'} = \frac{\alpha'_{\text{eff}}}{\alpha'}\, \hat{X}^{A'}.
\end{align}
With this scaling, the Polyakov action \eqref{eq:polyakov} can be rewritten as
\begin{equation}\begin{aligned}
	S_{\text{P}}=&~-\frac{1}{4\pi\alpha'_{\text{eff}}}\int d^{2}\sigma\sqrt{-\gamma}\gamma^{ab}
		\left(
			\partial_{a}X^{A}\partial_{b}X_{A}
			+
			\frac{\alpha'}{\alpha'_{\text{eff}}}
			\partial_{a}X^{A'}\partial_{b}X_{A'}
		\right)
	\\
	=&~-\frac{1}{4\pi\alpha'_{\text{eff}}}\int d^{2}\sigma\sqrt{-\gamma}\gamma^{ab}
		\left(
			\partial_{a}X^{A}\partial_{b}X_{A}
			+
			2\lambda_{a }^{A'}\partial_{b}X_{A'}
			-
			\frac{\alpha'_{\text{eff}}}{\alpha'}
			\lambda_{a }^{A'}\lambda_{b A' }
		\right)
		\,.
\end{aligned}\end{equation}
In the second line we have introduced auxiliary fields $\lambda_{a}^{A'}$, which can be integrated out to recover the first line. Taking the $\alpha' \to \infty$ limit, we obtain
\begin{align}\label{eq:magneticaction}
    S_{\text{M}} = -\frac{1}{4\pi \alpha'_{\text{eff}}} \int d^2\sigma \sqrt{-\gamma} \gamma^{a b} \left( \partial_a X^A \partial_b X_A + 2 \lambda_a^{A'} \partial_b X_{A'} \right)\,,
\end{align}
which we refer to as the \emph{magnetic} Carroll model.\footnote{A corresponding \emph{electric} Carroll model is obtained by a different scaling, $\hat{X}^{A} = X^A$, $\hat{X}^{A'} = \sqrt{\frac{\alpha'}{\alpha'_{\text{eff}}}} X^{A'}$. In this case, taking the $\alpha' \to \infty$ limit leads to 
$
    S_{\text{E}} = -\frac{1}{4 \pi \alpha'_{\text{eff}}} \int d^2\sigma \sqrt{-\gamma} \gamma^{ab} \partial_a X^{A'} \partial_b X_{A'}\,,
$
which corresponds to the Polyakov action for the transverse coordinate fields alone.} Upon adopting a conformal gauge for the worldsheet metric,
\begin{equation}\label{eq:confgauge}
	ds^{2}
	=
	-d\sigma^{+}d\sigma^{-}
	\,,
	\qquad
	\sigma^{\pm}=\tau\pm\sigma
	\,,
	\qquad
	\partial_{\pm}=\frac{1}{2}\left(\partial_\tau\pm\partial_\sigma\right)
	\,,
\end{equation}
we can write the magnetic action \eqref{eq:magneticaction} as
\begin{equation} \label{eq:magneticmodel}
	S_{\text{M}}
	=
	2 \, T_{\text{eff}} \, \int d^{2}\sigma
		\left(
			\partial_{+}X^{A}\partial_{-}X^{B}\eta_{AB}
			+
			\lambda_{+ }^{A'}\partial_{-}X_{A'}
			+
			\lambda_{- }^{A'}\partial_{+}X_{A'}
		\right)
		\,,
\end{equation}
where we have introduced the effective string tension $T_{\text{eff}} = 1/(2\pi \alpha'_{\text{eff}})$, and defined
\begin{align} \label{eq:lambda-conf}
    \lambda_\pm^{A'} = \frac{1}{2}\left( \lambda_\tau^{A'} \, \pm \, \lambda_\sigma^{A'} \right)\,.
\end{align}
In conformal gauge, the field equations of the magnetic Carroll model are
\begin{equation}\label{Eq:conformalMagEOM}
\begin{array}{rr}
    \partial_+ \partial_- X^A = 0\,,&
   \partial_{+} X^{A'} =0\,,\\ 
   \partial_{-} X^{A'} =0\,,& 
  \hspace{1cm}  \partial_+ \lambda_-^{A'} + \partial_- \lambda_+^{A'}  = 0\,,
\end{array}
\end{equation}
supplemented by constraints of the form
\begin{align} \label{eq:constraint-magnetic}
\partial_\pm X^A\partial_\pm X_A
+2\lambda_\pm^{A'}\partial_{\pm}X_{A'} =0\,.
\end{align}
It is immediately apparent that $X^{A'}$ is constant, {\it i.e.} there is no transverse motion in the magnetic Carroll model, and the constraints reduce to
\begin{align} \label{eq:reduced-constraint}
\partial_\pm X^A\partial_\pm X_A =0\,.
\end{align}
The dispersion relation of magnetic Carroll strings, 
\begin{align} \label{eq:spectral-magnetic}
    (P^0)^2 =(P^1)^2 \,,
\end{align}
follows from the worldsheet field equations and constraints in a straightforward way. To see this we consider null combinations of the longitudinal coordinate fields, $X^\pm=\frac{1}{\sqrt{2}}(X^0\pm X^1)$, 
and fix one of them by the usual lightcone gauge condition,
\begin{equation}\label{eq:lightconegauge}
X^{+}=x^{+}+\alpha^{\prime}_{\text{eff}}\,P^+ \tau\,. 
\end{equation}
This is manifestly a solution of the two-dimensional wave equation as required by \eqref{Eq:conformalMagEOM}. The other longitudinal field is a solution of the wave equation and a periodic function of $\sigma$, which results in a general mode expansion of the form
\begin{align} \label{eq:magnetic-sol-eom}
    X^{-} = x^{-} +  \alpha^{\prime}_{\text{eff}}\,P^{-} \,\tau + i\, \sqrt{\frac{\alpha^{\prime}_{\text{eff}}}{2}}\, \sum_{n \neq 0} \frac{1}{n} \left( \alpha_n^{-} e^{-in \sigma^-}+ \tilde{\alpha}_n^{-} e^{-in \sigma^+}\right)
    \,.
\end{align}
The longitudinal fields must also satisfy the constraints \eqref{eq:reduced-constraint}, which reduce to
\begin{align} \label{eq:lightcone-constraint}
P^+\partial_\pm X^- =0\,,
\end{align}
when the lightcone gauge condition \eqref{eq:lightconegauge} is satisfied. There are two cases to consider. If $P^+\neq 0$ then the constraints require $X^-$ to be a constant, which implies $P^-=0$. On the other hand, if $P^+=0$ then $X^-$ is unrestricted. In both cases the product of lightcone momenta vanishes, $P^+P^-=0$,
which is equivalent to \eqref{eq:spectral-magnetic}. 

We note that the energy of the magnetic Carroll string is independent of the transverse momentum.
The transverse momentum itself, however, remains unconstrained as it only appears in 
\begin{equation}
	P^{A'}
	=
	T_{\text{eff}}
	\int d\sigma
	\lambda^{A'}_{\tau}
	\,,
\end{equation}
where the zero mode of $\lambda^{A'}_{\tau}$ remains undetermined by the constraints.

Finally, if the spatial longitudinal coordinate is compactified on a circle of radius $R$,
\begin{equation}
X^{1}\sim X^{1}+2\pi R\,, 
\end{equation}	
then the energy takes discrete values $P^0=n/R$. 

\subsection{Canonical formulation}
In this subsection, we present the first-order formulation of the magnetic model to facilitate comparison with the Carroll string model found by Cardona, Gomis, and Pons (CGP) in \cite{Cardona:2016ytk}. 
The canonical momentum densities conjugate to the dynamical fields $X^{A}$ and $X^{A'}$ of the magnetic Carroll action \eqref{eq:magneticaction} are given by
\begin{align}
    \Pi_A = \frac{\delta \mathcal{L}_{\text{M}}}{\delta \dot{X}^A} = -T_{\text{eff}}\,  \sqrt{-\gamma} \gamma^{a \tau} \partial_a X_A \,, \hspace{1cm} \Pi_{A'} = \frac{\delta\mathcal{L}_{\text{M}}}{\delta \dot{X}^{A'}} = -T_{\text{eff}} \, \sqrt{-\gamma} \gamma^{a \tau} \lambda_{aA'}\,,
\end{align}
where the dot denotes a partial derivative with respect to $\tau$. The canonical conjugate momenta of the worldsheet metric $\gamma_{ab}$ and the auxiliary field $\lambda_{a A'}$ are identically zero and thus correspond to the primary constraints of the model. The canonical Hamilton density $\mathcal{H}_{\text{M}}=\Pi_\mu \dot{X}^{\mu}-\mathcal{L}_{\text{M}}$ can be written as a sum of a longitudinal and transverse part,
\begin{eqnarray} \label{eq:CanonicalSplit}
 \mathcal{H}_{\text{M}}^{\|} 
 &=& \Pi_A \partial_\tau X^{A} + \frac{1}{2} T_{\text{eff}} \sqrt{-\gamma} \gamma^{ab} \partial_a X^{A} \partial_b X_{A} \nonumber\\
 &=& \frac{\sqrt{-\gamma}}{2\, T_{\text{eff}}\,  \gamma_{\sigma \sigma}} \left( \Pi_A \Pi^A + T_{\text{eff}}^2 \, \partial_\sigma X_A \partial_\sigma X^A \right) + \frac{\gamma_{ \sigma\tau}}{\gamma_{\sigma \sigma}} \,\Pi_{A} \partial_\sigma X^{A}\,\,,\\
\mathcal{H}_{\text{M}}^\perp 
&=& \Pi_{A'} \partial_\tau X^{A'} +T_{\text{eff}} \sqrt{-\gamma} \gamma^{ab} \lambda_a^{A'} \partial_b X_{A'} \nonumber \\
&=& \frac{T_{\text{eff}}\, \sqrt{-\gamma}}{\gamma_{\sigma \sigma}}\lambda_{\sigma A'}\partial_{\sigma }X^{A'}+\frac{\gamma_{\sigma\tau}}{\gamma_{\sigma \sigma}}\,\Pi_{A'} \partial_\sigma X^{A'}\,.
\end{eqnarray}
We note that the longitudinal part has the standard form obtained from the Polyakov action. 
It follows that the Hamilton equations for the magnetic Carroll model are given by
\begin{eqnarray}
 \dot{\Pi}_{A} -  \partial_\sigma\left(\frac{\gamma_{\sigma\tau}}{\gamma_{\sigma\sigma}}  \Pi_{A}+\frac{T_{\text{eff}}\sqrt{-\gamma}}{\gamma_{\sigma\sigma}} \, \partial_\sigma X_{A} \right) & =& 0\,,\\ 
  \dot{X}^{A}  - \frac{\gamma_{\sigma\tau}}{\gamma_{\sigma\sigma}} \partial_\sigma X^{A} - \frac{\sqrt{-\gamma}}{T_{\text{eff}}\,\gamma_{\sigma\sigma}} \, \Pi_{B}\, \eta^{AB} & =& 0\,,\\
 \dot{\Pi}_{A'} -  \partial_{\sigma }\left( \frac{\gamma_{\sigma\tau}}{\gamma_{\sigma\sigma}}\, \dot{\Pi}_{A'} +\frac{T_{\text{eff}}\sqrt{-\gamma}}{\gamma_{\sigma\sigma}} \, \lambda_{\sigma A'}\right)  & =& 0\,,\\ 
 \dot X^{A'} - \frac{\gamma_{\sigma\tau}}{\gamma_{\sigma\sigma}}\, \partial_\sigma X^{A'} &=&0\,.
\end{eqnarray}
As expected, these equations reduce to \eqref{Eq:conformalMagEOM} in the conformal gauge \eqref{eq:confgauge}.

The action of the CGP Carroll string action in \cite{Cardona:2016ytk} is given by
\begin{align}\label{eq:CGPaction}
    S_{\text{CGP}} = \int d^2\sigma \left( \Pi \cdot \dot{X} - \mu \, \Pi \cdot X' - \frac{e}{2}\, \left( \Pi_A \Pi^A + T_{\text{eff}}^2\, \partial_\sigma X^{A'} \partial_\sigma X_{A'} \right) \right)\,,
\end{align}
where $\Pi \cdot \dot{X} \equiv \Pi_A  \partial_\tau X^{A} + \Pi_{A'} \partial_\tau X^{A'}$ and $\Pi \cdot X' \equiv \Pi^A \partial_\sigma X_A + \Pi^{A'} \partial_\sigma X_{A'}$. Here $\mu$ and $e$ are Lagrange multipliers enforcing the primary constraints.
The action of the magnetic Carroll model \eqref{eq:magneticaction} takes a similar form in first-order variables,
\begin{align} \label{eq:firstorder-mag}
    S_{\text{M}} = \int d^2\sigma &\left[\Pi \cdot \dot{X} - \frac{\gamma_{\sigma\tau}}{\gamma_{\sigma \sigma}} \Pi \cdot X' - \frac{\sqrt{-\gamma}}{2\, T_{\text{eff}} \, \gamma_{\sigma \sigma}}\left(\Pi_A \Pi^A + T_{\text{eff}}^2\left({X'}^{A} X'_{A} + 2\lambda_{\sigma}^{A'}  X'_{A'}\right) \right) \right]\,,
\end{align}
but there are significant differences between the two models. The longitudinal part of the magnetic Carroll action retains the standard Polyakov form while for the CGP action the longitudinal part $T_{\text{eff}} {X'}^{A}X'_A$ is absent and in the transverse fields of the magnetic Carroll model are constrained by the auxiliary fields $\lambda_\sigma^{A'}$. Notably, neither model has a term quadratic in transverse momenta  $\Pi^{A'}\Pi_{A'}$. 

The Hamilton equations governing the dynamics of the two models in a conformal gauge are listed in Table \ref{tab:different-hamilton}. For completeness, we also include the dynamical equations (also in a conformal gauge) of the chiral Carroll model studied in Section \ref{sec:chiralcarroll} below.
\begin{table}[ht!]
    \centering 
    \begin{tabular}{|l|l|l|l|l|} 
    \hline
         \textbf{Carroll string}  &\ \ $\dot{X}^{A} = \pdv{{H}}{\Pi_A}$ & $\dot{\Pi}^A = -\pdv{{H}}{X_A}$ & $\dot{X}^{A'} = \frac{\partial {H}}{\partial \Pi_{A'}}$ & $\dot{\Pi}^{A'} = -\pdv{{H}}{X_{A'}}$  
    \rule{0pt}{2.6ex} \rule[-1.5ex]{0pt}{0pt}      \\   \hline  \hline
       CGP \cite{Cardona:2016ytk} & $T_{\text{eff}} \dot{X}^A = \Pi^A$ & $\dot{\Pi}^A = 0$ & $\dot{X}^{A'} = 0$ & $\dot{\Pi}^{A'} = T_{\text{eff}} \,X^{\prime\prime A'}$  \rule{0pt}{2.4ex} \rule[-0.5ex]{0pt}{0pt} \\ \hline Magnetic model & $T_{\text{eff}}\dot{X}^A = \Pi^A$ & $\dot\Pi^A =T_{\text{eff}}\, X^{\prime\prime A}$ & $\dot{X}^{A'} = 0$ & $\dot\lambda_\tau^{A'} = \lambda_\sigma^{\prime A'}$  \rule{0pt}{2.4ex} \rule[-0.5ex]{0pt}{0pt} \\ \hline
       Chiral model & $T_{\text{eff}} \dot{X}^A = \Pi^A$ & $\dot\Pi^A =T_{\text{eff}}\, X^{\prime\prime A}$ & $\dot{X}^{A'} = X^{\prime A'}$ & $\dot\lambda_+^{A'} = \lambda_+^{\prime A'}$ \rule{0pt}{2.4ex} \rule[-0.5ex]{0pt}{0pt} \\ \hline
    \end{tabular}
    \caption{Dynamical equations for three different Carroll string models in conformal gauge.}
    \label{tab:different-hamilton}
\end{table}
The field equations place strong restrictions on the dynamics in all three models. In particular, there is no transverse motion at all allowed in the CGP and magnetic Carroll models. The chiral model is less constrained in this respect and supports left-moving transverse modes. Longitudinal momentum is constant in the CGP Carroll model but the magnetic and the chiral Carroll models do not have that restriction.

\subsection{Symmetries of the magnetic Carroll model}  
The magnetic Carroll action \eqref{eq:magneticaction} is invariant under  infinitesimal symmetry transformations of the form
\begin{subequations}\label{eq:symmetries2}
\begin{eqnarray}
	\delta X^{A} & =& K^{A} + \Lambda^{A}{}_{A'}X^{A'}+\Lambda^{A}{}_{B}X^{B}
	\,,
	\\
	\delta X^{A'} &=& K^{A'} + \Lambda^{A'}{}_{B'}X^{B'}
	\,, \\
 \delta \lambda_{a}^{A'} &=& -{\Lambda_{B'}}^{A'} \lambda_{a}^{B'} - {\Lambda_{A}}^{A'} \partial_a X^{A}
	\,,
\end{eqnarray}
\end{subequations}
where $K^{A}$ and $K^{A'}$ denote  longitudinal and transverse target space translations respectively, $\Lambda^{A}{}_{B}$ and $\Lambda^{A'}{}_{B'}$ denote a longitudinal boost and a transverse rotation\footnote{Here both parameters are skew-symmetric, {\it i.e.} $\Lambda_{AB}=-\Lambda_{BA}$ and $\Lambda_{A'B'}=-\Lambda_{B'A'}$.} respectively, and $\Lambda^{A}{}_{A'}$ denotes a Carroll boost from the longitudinal sector to the transverse sector. The corresponding Noether charges are 
\begin{equation}
\label{eq:magnetic-charges}\begin{aligned}
P_{A} &= \int d \sigma\, \Pi_{A}\,, \hspace{2.45cm} P_{A'} = \int d \sigma\, \Pi_{A'}\,, \\
{M_{AB}} &=  2\int d \sigma \, \Pi_{[A}X_{B]} \,, \hspace{1cm} 
M_{A'B'} =  2\int d \sigma \, \Pi_{[A'}X_{B']} \,, \\
{C_{AB'}} &= \int d\sigma \, \Pi_{A} X_{B'}\,.
\end{aligned}
\end{equation}
The charges form a closed algebra with the non-vanishing commutators given by
\begin{subequations}\label{eq:magnetic-charges_algebra}
\begin{eqnarray}
\left[ P_{A'} \,, {C_{AB'}} \right] & =& -\delta_{A'B'} P_{A}\,,\\
\left[ M_{AB}\,, C_{CA'}  \right] & =&2\eta_{C[B}\,C_{A]A'} \,, \\
\left[ {M_{A'B'}}\,, C_{AC'}  \right] & =&-2\delta_{C'[A'}\,C_{|A|B']} \,, \\
\left[ M_{A'B'}\,, P_{C'}\right] & = &2\delta_{C'[B'} P_{A']}\,,\\
\left[ M_{AB}\,, P_{C}\right] & = &2\eta_{C[B} P_{A]}\,,\\
\left[ M_{AB}\,,  M_{CD}\right] & =& 4\, \eta_{[C[B} M_{A]D]}  \,, \\
\left[ M_{A'B'}\,,  M_{C'D'}\right] & =& 4\,  \delta_{[C'[B'} M_{A']D']}\,.
\end{eqnarray}
\end{subequations}
This algebra was proposed in \cite{Cardona:2016ytk}, where it was obtained as a contraction of the Poincaré algebra and referred to as the \emph{string Carroll algebra}. We adopt the same terminology here.

The magnetic Carroll action \eqref{eq:magneticaction} is also invariant under an infinitesimal dilatation with scale parameter $\ell$ that acts only on the transverse sector,
\begin{align}\label{eq:dilation}
    \delta X^{A'} = \ell \, X^{A'} \,, \hspace{1cm} \delta \lambda_{a}^{A'} = -\ell \, \lambda_{a}^{A'}\,.
\end{align}
The corresponding Noether charge is 
\begin{align}\label{eq:dila}
    D = \frac{1}{2\pi \alpha'_{\text{eff}}}\int d\sigma\, \lambda_\tau^{A'} X_{A'} = \int d\sigma \, \Pi^{A'} X_{A'}\,,
\end{align}
and the non-vanishing commutators with the rest of the symmetry generators are
\begin{equation}
 \left[ D, \, P_{A'}\right]=P_{A'}\,, \hspace{1cm} \left[ D, \, C_{AA'} \right]=- C_{AA'}\,.
\end{equation}
This is reminiscent of the additional longitudinal dilatation symmetry found in non-relativistic string theory \cite{Bergshoeff:2019pij}, which gives rise to an additional gauge field in the gauging procedure for the string Newton-Cartan algebra. Here, the transverse dilatation symmetry \eqref{eq:dilation} is expected to be associated with an extra gauge field when gauging the string Carroll algebra \eqref{eq:magnetic-charges_algebra}.

The non-relativistic string in flat spacetime admits an infinite dimensional symmetry algebra, which contains the string Newton-Cartan algebra as a finite dimensional subalgebra \cite{Bergshoeff:2019pij,Batlle:2016iel}. A similar story plays out in the CGP Carroll string model \cite{Cardona:2016ytk} and also in the Carroll string models considered here. The magnetic and chiral Carroll models admit infinite dimensional symmetries, which contain as finite dimensional subalgebras the string Carroll algebra and the extended string Carroll algebra (defined in Section~\ref{sec:chiralcarrollsymmetries} below), respectively. As it turns out, the symmetries of the magnetic Carroll model are a subset of those of the chiral Carroll model and therefore we postpone the discussion of the infinite dimensional symmetries until we have introduced the chiral Carroll model in Section~\ref{sec:chiralcarroll}. 

\section{Chiral Carroll string model} \label{sec:chiralcarroll}
In this section, we consider a novel Carroll limit on target space. The construction involves the same asymmetric scaling of longitudinal and transverse coordinates as in the magnetic Carroll model but this time around an extra set of auxiliary fields is introduced that allows us to impose a worldsheet chirality on the transverse coordinate fields. The resulting model realises a non-trivial extension of string Carroll symmetry and the transverse dynamics turn out to be less constrained than in the magnetic Carroll model. 

\subsection{Relativistic origin of the chiral Carroll model}
Our starting point is once again the Polyakov action \eqref{eq:polyakov} with a rescaling of the coordinate fields as in  \eqref{eq:scaling} but now we use two sets of auxiliary fields $\lambda_{aA'}$ and $\chi_{aA'}$ to rewrite the action,\footnote{The equations of motion for $\lambda_a^{A'}$ and $\chi_a^{A'}$ implied by \eqref{eq:mainscaling} are covariantly given by:\begin{align}\label{eq:infootnote} \chi_a^{A'} = \left( \frac{\alpha'}{\alpha'_{\text{eff}}} \right)^2 \gamma_{ab} \Gamma^{bc}_+ \partial_c X^{A'}, \hspace*{1cm}  \lambda_a^{A'} = \left(\frac{\alpha'}{\alpha'_{\text{eff}}} \right) \gamma_{ab} \Gamma_-^{bc} \partial_c X^{A'}\end{align} }
\begin{equation}\label{eq:mainscaling}
\begin{aligned}
	S_{\text{P}}=&~-\frac{1}{4\pi\alpha_{\text{eff}}'}\int d^{2}\sigma\sqrt{-\gamma}\gamma^{ab}
		\left[
			\partial_{a}X^{A}\partial_{b}X_{A}
			+
			\frac{\alpha'}{\alpha'_{\text{eff}}}
			\partial_{a}X^{A'}\partial_{b}X_{A'}
		\right]
	\\
	=&~-\frac{1}{4\pi\alpha_{\text{eff}}'}\int d^{2}\sigma\sqrt{-\gamma}
		\left[
			\gamma^{ab}\partial_{a}X^{A}\partial_{b}X_{A}
			+
			2\,\Gamma_{+}^{ab}\lambda_{a A'}\partial_{b}X^{A'}
			\right.
			\\&
			\left.
			\qquad\qquad\qquad\qquad\qquad
			+\, 
			\frac{2\alpha'_{\text{eff}}}{\alpha'}\, \Gamma_{-}^{ab}\chi_{a A'}\partial_{b}X^{A'}
			-
		2\left(\frac{\alpha_{\text{eff}}'}{\alpha'}\right)^{2}\gamma^{ab}\lambda_{a}^{A'}\chi_{bA'}
		\right]
		\,,
\end{aligned}\end{equation}
where we have introduced the chiral projection operators 
\begin{equation} \label{eq:chiral-projectors}
\Gamma^{ab}_{\pm}=\frac12\left(\gamma^{ab}\pm\epsilon^{ab}\right)
	\,,
\end{equation}
with $\epsilon^{ab}$ the Levi-Civita tensor. We employ the conventions $\epsilon^{ab}={\overline{\epsilon}}^{ab}/\sqrt{-\gamma}\,,$ $\epsilon_{ab} =	\overline{\epsilon}_{ab} \sqrt{-\gamma}$, where $\overline{\epsilon}^{ab}$ denotes the Levi-Civita symbol, $\overline{\epsilon}^{01}=+1=-\overline{\epsilon}^{10}$.
We can now readily take the limit $\alpha'\to\infty$ to obtain a \emph{chiral Carroll string model},
\begin{align} \label{eq:chiral-Carroll-action}
        S_{\text{C}} = -\frac{1}{4\pi \alpha'_{\text{eff}}} \int d^2\sigma \sqrt{-\gamma} \left( \gamma^{ab} \partial_{a} X^{A} \partial_{b} X_{A} \; +\; 2\, \Gamma_{+}^{ab} \lambda_a^{A'} \partial_{b} X_{A'} \right)\,.
\end{align}
Upon adopting the conformal gauge (\ref{eq:confgauge}) the chiral Carroll action reduces to
\begin{equation}\label{eq:chiralmodel}
	S_{\text{C}}
	=
	\frac{1}{\pi \alpha'_{\text{eff}}}\int d^{2}\sigma
	\left(
		\partial_{+}X^{A}\partial_{-}X_{A}
		+
		\lambda_{+}^{A'}\partial_{-}X_{A'}
	\right)
	\,.
\end{equation}
Compared to the magnetic model \eqref{eq:magneticmodel} we have effectively projected out one worldsheet chirality of the transverse fields $X^{A'}$ and $\lambda_{aA'}$.\footnote{We have chosen to retain the left-moving fields $\lambda_{+A'}$. We could of course just as well have projected onto the other chirality.} The new auxiliary fields $\chi_{aA'}$ do not appear in the final action in the $\alpha'\to\infty$ limit but they nevertheless play an important role in implementing the projection onto a single worldsheet chirality.  In conformal gauge the equations of motion obtained from the chiral Carroll action reduce to
\begin{equation} \label{eq:eomchiralcarroll}
	\partial_{+}\partial_{-}X^{A}=\partial_{-}\lambda_{+}^{A'} = \partial_{-}X^{A'}=0
	\,.
\end{equation}
This is a subset of the field equations of the magnetic model \eqref{Eq:conformalMagEOM} and therefore the dynamics is now less restrictive than before. In particular, 
the transverse fields $X^{A'}$ can have non-trivial dependence on the $\sigma^{+}$ direction in the chiral model, in contrast to the magnetic model where the transverse fields are only allowed to take constant values. The energy momentum tensor constraints are now given by 
\begin{align}
    \begin{split}
     0 &= \partial_+ X^{A} \partial_+ X_{A} + \lambda_+^{A'}\partial_+ X_{A'}\,, \\
     0 &= \partial_- X^{A} \partial_- X_{A} \,.
    \end{split}
\end{align}
By implementing the lightcone gauge condition 
\begin{align}
    X^+ = x^+ + \alpha'_{\text{eff}} P^+ \tau \,,
\end{align}
and inserting it into the constraints we find that
\begin{align}
\begin{split} \label{eq:lq-chiral-carroll}
    \partial_+ X^- &= \frac{1}{\alpha'_{\text{eff}}P^+} \lambda_+^{A'} \partial_+ X_{A'}\,, \\
    \partial_- X^- &= 0\,. 
\end{split}
\end{align}
From the latter it follows that $X^-$ is also chiral and hence the mode expansion for $X^-$ can be written in the following form
\begin{align}
    X^-(\sigma^-,\sigma^+) = X^{-}_L(\sigma^+) = x^- + \frac{1}{2}\alpha'_{\text{eff}} P^- \sigma^+ + i\sqrt{\frac{\alpha'_{\text{eff}}}{2}} \sum_{n \neq 0} \frac{1}{n} \alpha_n^{-} e^{-in\sigma^+}\,.
\end{align}
However, since $X^-$ is periodic in $\sigma$ this then implies that $P^- = 0$ i.e. that $E = P_1$. Therefore, although we have transverse excitations, they do not contribute to the energy. Now consider the transverse fields which are left-moving and periodic in $\sigma$,
\begin{align}
    \begin{split}
    X^{A'}(\sigma^+) &= x^{A'} + i \sqrt{\frac{\alpha'_{\text{eff}}}{2}} \sum_{n \neq 0} \frac{1}{n} \alpha_n^{A'} e^{-in\sigma^+},  \\
    \lambda_+^{A'}(\sigma^+) &= \sqrt{\frac{\alpha'_{\text{eff}}}{2}}\sum_n \lambda_n^{A'} e^{-in\sigma^+},
\end{split}
\end{align}
where we note in particular that the transverse momenta $P^{A'}$ are related to the zero modes of the auxiliary fields $\lambda^{A'}$ via
\begin{align}
    P^{A'} = \int_0^{2\pi} d\sigma \Pi^{A'} = \frac{1}{\sqrt{2 \alpha'_{\text{eff}}}} \lambda_0^{A'}.
\end{align}
As usual, the longitudinal $\alpha_n^-$ are obtained by inserting the mode expansions into the constraint in \eqref{eq:lq-chiral-carroll}. We find that for $n \neq 0$ we have
\begin{align}
    \alpha_n^- = \frac{1}{P^+\sqrt{2 \alpha'_{\text{eff}}}} \sum_{m} \lambda_m^{A'}\alpha_{n-m}^{B'}\delta_{A'B'},
\end{align}
where we have additionally defined $\alpha_0^{A'} = 0$ and $\alpha_0^- = 0$ for notational convenience. The zero mode of the constraint then implies that
\begin{align}
    \sum_{n} \lambda^{A'}_n \alpha^{B'}_{-n}\delta_{A'B'} = 0\,,
\end{align}
which says the total left-moving excitation level must be zero after all.

In order to have non-zero string excitations one needs to compactify a transverse direction, say $A'=D-1$, such that
\begin{equation}
	X^{D-1}(\sigma+2\pi)
	=
	X^{D-1}(\sigma )
	+
	2\pi w \mathcal{R}
	\,.
\end{equation}
where $w$ is the winding number.
We now have a zero mode and a quantised momentum in the compact direction,
\begin{equation}
	\alpha_0^{D-1} 
	= 
	\sqrt{\frac{2}{\alpha^\prime_{\text{eff}}}} w \mathcal{R} \,,
	\qquad 
	P^{D-1}=\frac{n}{\mathcal{R}} \,,
\end{equation}
and the two constraint equations in \eqref{eq:lq-chiral-carroll} become
\begin{equation}\label{eq:final_disp}
	E
	=
	P_{1}
	\,,
	\qquad
	\sum_{n>0} \lambda^{A'}_n \alpha^{B'}_{-n}\delta_{A'B'}
	=
	-w n
	\,,
\end{equation}
where we can interpret the second equation as a level matching condition that determines
the left-moving excitation level in terms of the winding number.

\subsection{Canonical formulation}
The canonical conjugate momenta for the longitudinal and transverse worldsheet fields of the chiral Carroll action \eqref{eq:chiral-Carroll-action} are given by  
\begin{align}
    \Pi_A = \frac{\delta \mathcal{L}_{\text{C}}}{\delta \dot{X}^A} = -T_{\text{eff}}\,  \sqrt{-\gamma} \gamma^{a \tau} \partial_a X_{A}\,, \hspace{1cm}
    \Pi_{A'} = \frac{\delta \mathcal{L}_{\text{C}}}{\delta \dot{X}^{A'}} = -T_{\text{eff}}\sqrt{-\gamma} \, \Gamma_+^{a \tau } \lambda_{a A'}\,.
\end{align}
The canonical Hamilton density $\mathcal{H}_{\text{M}}=\Pi_\mu \dot{X}^{\mu}-\mathcal{L}_{\text{M}}$ can be written as a sum of a longitudinal and transverse part,
\begin{eqnarray} \label{eq:ChiralCanonicalSplit}
\mathcal{H}_{\text{C}}^{\|} 
 &=& \Pi_A \partial_\tau X^{A} + \frac{1}{2} T_{\text{eff}} \sqrt{-\gamma} \gamma^{ab} \partial_a X^{A} \partial_b X_{A} \nonumber\\
 &=& \frac{\sqrt{-\gamma}}{2\, T_{\text{eff}}\,  \gamma_{\sigma \sigma}} \left( \Pi_A \Pi^A + T_{\text{eff}}^2 \, \partial_\sigma X_A \partial_\sigma X^A \right) + \frac{\gamma_{ \sigma\tau}}{\gamma_{\sigma \sigma}} \,\Pi_{A} \partial_\sigma X^{A}\,,\\
\mathcal{H}_{\text{C}}^\perp 
&=& \Pi_{A'} \partial_\tau X^{A'} +T_{\text{eff}} \sqrt{-\gamma} \gamma^{ab} \lambda_a^{A'} \partial_b X_{A'} \nonumber \\
&=& \frac{(\gamma_{\sigma\tau}+\epsilon_{\sigma \tau})}{\gamma_{\sigma \sigma}}\,\Pi_{A'} \partial_\sigma X^{A'}\,.
\end{eqnarray}
The longitudinal part is given by the same Polyakov expression as we found for the magnetic model in \eqref{eq:CanonicalSplit}. The transverse part  differs from the magnetic model in that it does not involve the auxiliary fields at all and incorporates chirality through the anti-symmetric Levi-Civita tensor $\epsilon_{ab}$. The Hamilton equations for the chiral model are 
\begin{eqnarray}
 \dot{\Pi}_{A} -  \partial_\sigma\left(\frac{\gamma_{\sigma\tau}}{\gamma_{\sigma\sigma}}  \Pi_{A}+\frac{T_{\text{eff}}\sqrt{-\gamma}}{\gamma_{\sigma\sigma}} \, \partial_\sigma X_{A} \right) & =& 0\,,\\ 
  \dot{X}^{A}  - \frac{\gamma_{\sigma\tau}}{\gamma_{\sigma\sigma}} \partial_\sigma X^{A} - \frac{\sqrt{-\gamma}}{T_{\text{eff}}\,\gamma_{\sigma\sigma}} \, \Pi_{B}\, \eta^{AB} & =& 0\,,\\
 \dot{\Pi}_{A'} -  \partial_{\sigma }\left( \frac{(\gamma_{\sigma\tau}+\epsilon_{\sigma\tau})}{\gamma_{\sigma\sigma}}\,  \dot{\Pi}_{A'} \right) & =& 0\,,\\ 
 \dot X^{A'} - \frac{(\gamma_{\sigma\tau}+\epsilon_{\sigma\tau})}{\gamma_{\sigma\sigma}}\, X^{\prime A'} &=&0\,.
\end{eqnarray}
In the conformal gauge \eqref{eq:confgauge}, these Hamilton equations reproduce \eqref{eq:eomchiralcarroll}
and they are also listed in Table~\ref{tab:different-hamilton} above for comparison with the CGP and magnetic Carroll models. The field equations of the chiral Carroll model only require the transverse fields $X^{A'}$ and their conjugate momenta to be constant along the $\sigma^{-}$ worldsheet direction. Therefore, unlike the CGP and magnetic Carroll  models, the chiral model allows for transverse excitations moving in one direction along the string. 

\subsection{Symmetries of the chiral Carroll model} 
\label{sec:chiralcarrollsymmetries}
The chiral Carroll action \eqref{eq:chiral-Carroll-action} is invariant under infinitesimal symmetry transformations on the target space of the form
\begin{equation}\label{eq:symmetries3}\begin{aligned}
	\delta X^{A}=~&\Lambda^{A}_{\;\;\;A'}X^{A'}+\Lambda^{A}{_B}X^{B}+K^{A}
	\,,
	\\
	\delta X^{A'}=~&\Lambda^{A'}_{\;\;\;B'}X^{A'}+K^{A'}
	\,,
	\\
	\delta \lambda_{a A'}=~& -\Lambda^{B'}{}_{A'} \lambda_{aB'} -\Lambda^{A}_{\;\;\;A'}\partial_{a}X_{A}
	\,,
\end{aligned}\end{equation}
where $\{K^{A}, K^{A'},\Lambda^{A}{}_{B},\Lambda^{A'}{}_{B'},\Lambda^{A}{}_{A'}\}$ denote respectively  longitudinal and transverse translations, longitudinal and transverse rotations, and Carroll string boosts. The corresponding Noether charges are given by
\begin{equation}
\label{eq:chiral-charges}\begin{aligned}
P_{A} &= \int d \sigma\, \Pi_{A}\,, \hspace{2.45cm} P_{A'} = \int d \sigma\, \Pi_{A'}\,, \\
{M_{AB}} &=  2\int d \sigma \, \Pi_{[A}X_{B]} \,, \hspace{1cm} M_{A'B'} =  2\int d \sigma \, \Pi_{[A'}X_{B']} \,, \\
{C_{AB'}} &= \int d\sigma \left( \Pi_{A} X_{B'} - T_{\text{eff}}\, X_{A}\partial_\sigma X_{B'}\right)\,,
\end{aligned}
\end{equation}
where we note that the extra term in the Carroll string boosts $C_{AB'}$ compared with \eqref{eq:magnetic-charges} arises from a total derivative in the symmetry invariance of the action. The non-vanishing commutators are 
\begin{subequations}
\label{eq:chiral-algebra1}
\begin{eqnarray}
    \comm{Z_{A'B'}}{P_{C'}} & =& 2\delta_{C'[A'}Z_{B']} \\
    \comm{M_{A'B'}}{Z_{C'}} & =& 2 \delta_{C'[A'}Z_{B']}, \\
    \comm{M_{A'B'}}{Z_{C'D'}} & =& 4 \delta_{[C'[A'}Z_{B']D']}, \\
    \comm{M_{AB}}{P_C} & =& 2\eta_{C[B}P_{A]}, \\
   \comm{C_{AB'}}{P_C} & =&- \eta_{CA}Z_{B'}, \\
   \comm{C_{AB'}}{P_{C'}} & =& \delta_{C'B'} P_A, \\
   \comm{M_{A'B'}}{P_{C'}} & =& -2\delta_{C'[A'}P_{B']}, \\
   \comm{M_{AB}}{M_{CD}} & =& 4\eta_{[C[A}M_{B]D]} \\
   \comm{M_{A'B'}}{M_{C'D'}} & =& 4\delta_{[C'[A'}M_{B']D']}  \\
   \comm{M_{AB}}{C_{CD'}} & =&- 2\eta_{C[A}C_{B]D'}, \\
   \comm{M_{A'B'}}{C_{CD'}} & =& -2 C_{C[A'} \delta_{B']D'}, \\
   \comm{C_{AB'}}{C_{CD'}} &= & \eta_{CA} Z_{B'D'},
   \end{eqnarray}
\end{subequations}
where we have defined additional charges,  
\begin{equation}\label{eq:extensions1}
Z_{A'} = T_{\text{eff}} \, \int d\sigma\, \partial_\sigma X_{A'}\,, \hspace{1.5cm} Z_{A' B'}= Z_{[A'B']} = 2\, T_{\text{eff}}\,\int d\sigma\,  X_{[A'}\, \partial_\sigma X_{B']} \,.
\end{equation}
With these extensions included, we have a closed algebra that we refer to as the \emph{extended string Carroll algebra}. This algebra can also be derived from an $\dot{\text{I}}$n\"on\"u-Wigner contraction of the coadjoint Poincaré algebra as we will explicitly show in subsection \ref{sec:inonu} below.

Finally, we mention that the algebra \eqref{eq:chiral-algebra1} can be further enlarged by an emergent scaling symmetry $D$ as in \eqref{eq:dila}.
This adds the following non-vanishing commutators
\begin{equation}
\comm{D}{ P_{A'}} \, = \, P_{A'}\,, \hspace{0.43cm}  \comm{D}{ Z_{A'}} \, = \,- Z_{A'}\,, \hspace{0.43cm} \comm{D}{ C_{AA'}} \, = \, -C_{AA'}\,, \hspace{0.43cm} \comm{D}{ Z_{A'B'}} = -2Z_{A'B'}\,.
\end{equation}
In the next subsection, we will see how the scaling symmetry arises from an infinite dimensional extension of \eqref{eq:chiral-algebra1}.

\subsection{Infinite dimensional symmetries of the chiral Carroll action} \label{sec:chiralcarroll-infsym}
The chiral Carroll action \eqref{eq:chiral-Carroll-action} is in fact invariant under an infinite dimensional symmetry,\footnote{These symmetry transformations can be derived through the conservation of a generator of canonical symmetry transformations $G$, given by \begin{equation}
  G= \int d\sigma \left( \xi^A \Pi_A +  \xi^{A'} \Pi_{A'} +\psi_{a}^{A'}\Pi_{a A'}+ \Lambda \pi_e+\gamma\pi_\mu \right)  
\end{equation} such that $\delta X^{A}=\{ G,X^{A}\} = \xi^{A}, \delta X^{A'}=\{ G,X^{A'}\} = \xi^{A'}$, and $\delta \lambda_{a}^{A'}=\{ G,\lambda_{a}^{A'}\} = \psi^{A'}_{a}$, where the functions $\xi^{A}, \xi^{A'}, \psi^{A'}_{a}, \Lambda$, and $\gamma$ satisfy differential equations obtained from $dG/d\tau=0$. }
\begin{subequations} \label{eq:chiral-generalized-symmetries}
\begin{eqnarray}
    \delta X^{A} &=& f^A(X^{A'}) + \Omega^{A}{}_{B}(X^{A'}) X^{B}\,, \\
    \delta X^{A'} &=& \xi^{A'}(X^{B'})\,, \\
    \delta \lambda_{aA'} &=& -\partial_a X_A \partial_{A'} f^{A} - \partial_a X_A \partial_{A'} \Omega^{A}{}_{B}  X^{B} -\lambda_{a B'} \partial_{A'} \xi^{B'}\,,
\end{eqnarray}
\end{subequations}
where $f^{A}, \Omega^{A}{}_{B}$ and $\xi^{A'}$ are arbitrary functions that depend only on the transverse coordinates $X^{A'}$ and $\Omega^{A}{}_{B}$ is anti-symmetric, $\Omega_{AB} = -\Omega_{BA}$. In the specific case of
\begin{align}
    f^A(X^{A'}) = K^A + \Lambda^{A}{}_{A'} X^{A'}\,, \hspace{0.5cm} \Omega^{A}{}_{B}(X^{B'}) = \Lambda^{A}{}_{B}\,, \hspace{0.5cm} \xi^{A'}(X^{B'}) = K^{A'} + \Lambda^{A'}{}_{B'}X^{B'}\,,
\end{align}
we recover the target space symmetries \eqref{eq:symmetries3} discussed above. Note that under the infinitesimal transformation $\delta X^{A} = f^{A}(X^{A'})$ the action is quasi-symmetric since under such a transformation we obtain a total derivative term,
\begin{align}
    \delta S_{\text{C}} = \int d^2\sigma \, \partial_a\left( T_{\text{eff}} \sqrt{-\gamma} \, \epsilon^{ab}\,  \partial_{A'} f^A X_A \partial_b X^{A'} \right)\,,
\end{align}
which gives an additional contribution to the conserved charges associated with $f^A$. Now consider expanding the functions $f^A, \Omega^{A}{}_{B}$ and $\xi^{A'}$ in a Taylor series of the transverse coordinates,
\begin{subequations} \label{eq:mag-series-expansion}
\begin{align}
    f^A(X^{A'}) &= a^A + a^{A}{}_{A'}X^{A'} + a^{A}{}_{A'B'}X^{A'}X^{B'} + \ldots  =\   a^{A}{}_{A'_1 A'_2 \cdots A'_n} X^{A'_1 A'_2 \cdots A'_n} , \\
   \Omega^{A}{}_{B}(X^{A'}) &= \omega^{A}{}_{B} + \omega^{A}{}_{B A'}X^{A'} + \omega^{A}{}_{B A'B'}X^{A'}X^{B'} + \ldots=  \omega^{A}{}_{B A'_1 A'_2 \cdots A'_n} X^{A'_1 A'_2 \ldots A'_n} , \\
   \xi^{A'}(X^{B'}) &= \varphi^{A'} + \varphi^{A'}{}_{B'}X^{B'} + \varphi^{A'}{}_{B' C'}X^{B'}X^{C'} + \ldots = \varphi^{A'}{}_{A'_1 A'_2 \cdots A'_n}X^{A'_1 A'_2 \ldots A'_n}  ,
\end{align}
\end{subequations}
where we have introduced the short hand notation $X^{A'_1 A'_2 \cdots A'_n} = X^{A'_1} X^{A'_2} \cdots X^{A'_n}$. For each of the functions $f^A, \Omega^{A}{}_{B}$ and $\xi^{A'}$, there is an infinite tower of conserved charges given by
\begin{subequations}\label{eq:generalized-chiral-charges}
\begin{eqnarray}
\xi^{A'}: \hspace{0.82cm}     K^{(n)}_{A'A'_1 A'_2 \cdots A'_n} &=& \int d\sigma \, \Pi_{A'} X_{A'_1 A'_2 \cdots A'_n}\,, \\
f^{A}: \hspace{1cm}    L^{(n)}_{A A'_1 A'_2 \cdots A'_n} &=& \int d\sigma \, \left( \Pi_{A} X_{A'_1 A'_2 \cdots A'_n} - T_{\text{eff}} \, X_A \partial_\sigma \left( X_{A'_1 \cdots A'_n} \right) \right)\,, \\
\Omega^{A}{}_{B}: \hspace{0.72cm}    M^{(n)}_{AB A'_1 A'_2 \cdots A'_n} &=& 2\int d\sigma \, \Pi_{[A}X_{B]} X_{A'_1 A'_2 \cdots A'_n}\,.
\end{eqnarray}
\end{subequations}
where the index $n \geq 0$ denotes the number of additional transverse indices. For instance, we recover the charges in \eqref{eq:chiral-charges} as follows,
\begin{align}
    P_{A'} = K_{A'}^{(0)}\,, \hspace{0.5cm} P_{A} = L_A^{(0)}\,, \hspace{0.5cm} M_{AB} = M_{AB}^{(0)}\,, \hspace{0.5cm} C_{AB'} = L_{A B'}^{(1)}\,, \hspace{0.5cm} M_{A'B'} = 2K_{[A' B']}^{(1)}\,.
\end{align} The infinite tower of charges in \eqref{eq:generalized-chiral-charges} satisfies an infinite dimensional Lie algebra,
\begin{subequations} \label{eq:chiral-generalized-algebra}
\begin{align}
    \comm{K_{A' A'_1 \cdots A'_n}^{(n)}}{K_{B' B'_1 \cdots B'_m}^{(m)}} &= \sum_{k=1}^{n} \delta_{B' A'_k} K_{A' A'_1 \cdots A'_{k-1} A'_{k+1} \cdots A'_n B'_1 \cdots B'_m}^{(n+m-1)} \\&\hspace{1cm} - \sum_{k=1}^{m} \delta_{A' B'_k} K_{B' A'_1 \cdots A'_n B'_1 \cdots B'_{k-1} B'_{k+1} \cdots B'_m}^{(n+m-1)}\,,  \\
    \comm{K^{(n)}_{A' A'_1 \cdots A'_n}}{L_{B B'_1 \cdots B'_m}^{(m)}} &= -\sum_{k=1}^{m} \delta_{A' B'_k} L^{(m+n-1)}_{B A'_1 \cdots A'_n B'_1 \cdots B'_{k-1} B'_{k+1} \cdots B'_m}\,, \\
    \comm{K_{A' A'_1 \cdots A'_n}^{(n)}}{M_{BC B'_1 \cdots B'_m}^{(m)}} &= - \sum_{k=1}^{m} \delta_{A' B'_k} M_{BC A'_1 \cdots A'_n B'_1 \cdots B'_{k-1} B'_{k+1}\cdots B'_m}^{(n+m-1)}\,, \\
    \comm{L_{A A'_1 \cdots A'_n}^{(n)}}{L_{B B'_1 \cdots B'_m}^{(m)}} &= \eta_{AB}\, Z^{(n,m)}_{A'_1 \cdots A'_n B'_1 \cdots B'_m}\,, \\
    \comm{L_{A A'_1 \cdots A'_n}^{(n)}}{M_{BC B'_1 \cdots B'_m}^{(m)}} &= 2\eta_{A[B} L^{(n+m)}_{C]A'_1 \cdots A'_n B'_1 \cdots B'_m} 
    \,, \\
    \comm{M_{AB A'_1 \cdots A'_n}^{(n)}}{M_{CD B'_1 \cdots B'm}^{(m)}} &= 4\eta_{[C[B} M^{(n+m)}_{A]D] A'_1 \cdots A'_n B'_1 \cdots B'_m}\,, 
\end{align}
\end{subequations}
where we have again introduced a set of extension generators,
\begin{align}\label{theextensions}
    Z^{(n,m)}_{A'_1 \cdots A'_n B'_1 \cdots B'_m} &= T_{\text{eff}} \int d\sigma \left( X_{A'_1 \cdots A'_n} \partial_\sigma \left( X_{B'_1 \cdots B'_m} \right) - X_{B'_1 \cdots B'_n} \partial_\sigma \left( X_{A'_1 \cdots A'_m} \right) \right)\,.
\end{align}
In the specific case of
\begin{align}
    Z^{(0,1)}_{A'} &= T_{\text{eff}} \int d\sigma \partial_\sigma X_{A'}\,, \hspace{1cm} Z^{(1,1)}_{A'B'} = T_{\text{eff}} \int d\sigma X_{[A'}\partial_\sigma X_{B']}\,,
\end{align}
 we recover the extensions defined in \eqref{eq:extensions1}. The only additional non-vanishing commutators that arise from adding these extensions are given by
\begin{align}
    \comm{Z^{(n,m)}_{A'_1 \cdots A'_n B'_1 \cdots B'_m}}{K^{(\ell)}_{C' C'_1 \cdots C'_\ell}} &= \sum_{k=1}^{m}\delta_{C'B^{\prime}_{k}}\, Z_{B'_1 \cdots B'_{k-1} B'_{k+1} \cdots B'_m C_1'\cdots C_l' A_1'\cdots A_n'}^{(m-1,l+n)} \\ &\hspace{1cm}
    + \sum_{k=1}^{n}\delta_{C'A^{\prime}_{k}}\, Z_{A_1'\cdots A'_{k-1} A'_{k+1} \cdots A_n'C_1'\cdots C_l' B_1'\cdots B_m'}^{(n-1,l+m)} \,.
\end{align}
The magnetic Carroll model \eqref{eq:magneticaction} also has an infinite dimensional symmetry algebra which is simply the subalgebra of \eqref{eq:chiral-generalized-algebra} obtained by setting the extensions \eqref{theextensions} to zero. 

The symmetry algebras of both the chiral and magnetic Carroll models include the following generators,
\begin{subequations} \label{eq:conformal-charges}
\begin{align}
    D &\equiv \delta^{A'B'} K^{(1)}_{A'B'}   = \int d\sigma \, \Pi_{A'}X^{A'}\,, \\
    K_{A'} & \equiv 2\delta^{B'C'} K_{B'C'A'}^{(2)} -  \delta^{B' C'} K^{(2)}_{A'B'C'} = \int d\sigma\,  \left( 2 \Pi_{B'}X^{B'} X_{A'} - \Pi_{A'}X^{B'}X_{B'} \right)\,.
\end{align}
\end{subequations}
The former corresponds to spatial dilatations and the latter corresponds to spatial special conformal transformations also found for the CGP Carroll string \cite{Cardona:2016ytk}. Furthermore, we can construct a longitudinal conformal transformation generator
\begin{align} \label{eq:conf-long}
    K_{A} = \delta^{A'B'}L^{(2)}_{A\, A'B'} = \int d\sigma \, \left( \Pi_{A} X_{A'} X^{A'} - T_{\text{eff}} X_A \partial_\sigma \left( X^{A'} X_{A'} \right) \right)\,.
\end{align}
The three generators in \eqref{eq:conformal-charges} and \eqref{eq:conf-long} are somewhat similar to the generators of the conformal Carroll algebra given by \cite{Basu:2018dub,Bagchi:2019xfx},
\begin{subequations} 
\begin{align}
    \tilde{D} &= \int d\sigma \left(\Pi^{A}X_{A} + \Pi^{A'}X_{A'} \right)\,, \\ \tilde{K}_{A} &= \int d\sigma X^{A'}X_{A'} \Pi_{A}\,, \\ \tilde{K}_{A'} &= \int d\sigma \left( 2X_{A'}(X^{A}\Pi_A + X^{B'}\Pi_{B'}) - (X^{B'}X_{B'}) \Pi_{A'} \right)\,.
\end{align}
\end{subequations}
but there are important differences. In particular, the longitudinal parts of $\tilde{D}$ and $\tilde{K}_{A'}$ are absent from our spatial special conformal transformations in equation \eqref{eq:conformal-charges} and we have an additional term associated with the quasi-symmetry in comparing $\tilde{K}_{A}$ with \eqref{eq:conf-long}.

\subsection{Extended Carroll algebra from the coadjoint Poincaré algebra}\label{sec:inonu}
In this section, we derive the extended string Carroll algebra \eqref{eq:chiral-algebra1} through an $\dot{\text{I}}$n\"on\"u-Wigner contraction of the coadjoint Poincar\'{e} algebra. The procedure is as follows: Let us consider 
an extension of the Poincaré algebra enlarged by the antisymmetric two-tensor $\hat{Z}_{\mu \nu}$ and the vector $\hat{Z}_{\mu}$ (see, for example, \cite{Barducci:2019jhj}), which consists of the following non-vanishing relativistic brackets\footnote{The hatted generators will be the ones associated to the relativistic algebra.}
\begin{align}
    \comm{\hat{M}_{\mu \nu}}{\hat{M}_{\rho \sigma}} &= 4\eta_{[\rho[\mu}\hat{M}_{\nu]\sigma]}\,, \\
    \comm{\hat{M}_{\mu \nu}}{\hat{P}_\rho} &= 2\eta_{\rho [\mu} \hat{P}_{\nu]}, \\
    \comm{\hat{M}_{\mu \nu}}{\hat{Z}_\rho} &= 2\eta_{\rho [\mu} \hat{Z}_{\nu]}, \\
    \comm{\hat{M}_{\mu \nu}}{ \hat{Z}_{\rho \sigma}} &= 4\eta_{[\rho[\mu}\hat{Z}_{\nu]\sigma]}\,, \\
    \comm{\hat{Z}_{\mu \nu}}{ \hat{P}_{\rho}} &= 2\eta_{\rho [\mu} \hat{Z}_{\nu]}\,,
    \end{align}
along with the following vanishing brackets
\begin{align}
    \comm{\hat{P}_\mu}{ \hat{P}_\nu } = \comm{ \hat{Z}_\mu}{ \hat{P}_\nu} = \comm{ \hat{Z}_\mu}{\hat{Z}_\nu} = \comm{\hat{Z}_{\mu\nu}}{ \hat{Z}_\rho} = \comm{\hat{Z}_{\mu \nu}}{\hat{Z}_{\rho \sigma}} =  0\,.
\end{align}
In order to obtain the extended string Carroll algebra from the algebra above, we scale the generators in the following manner
\begin{align}
    \hat{P}_{A} &= \omega \, P_{A}\,, \hspace{4.5cm} \hat{Z}_A = \omega \,Z_{A}, \\
   \hat{P}_{A'} &= P_{A'} + \frac{\omega^2}{2} \, Z_{A'}\,, \hspace{3cm}  \hat{Z}_{A'} = \omega \, Z_{A'}\,, \\
    \hat{M}_{AB} &= M_{AB}\,, \hspace{4.3cm} \hat{Z}_{AB} = Z_{AB}\,, \\
    \hat{M}_{A'B'} &= M_{A'B'}\,,\hspace{3.9cm} \hat{Z}_{A'B'} = \omega \, Z_{A'B'} \,, \\
   \hat{M}_{AB'} &= \omega \, C_{AB'} - \frac{\omega^2}{2}Z_{AB'}\,, \hspace{2cm} 
   \hat{Z}_{AB'} = \omega \, Z_{AB'}\,,
\end{align}
 where we introduced longitudinal indices $A = 0,1$ and transverse indices $A' = 2, \ldots, D-1$, and $\omega$ a dimensionless parameter. These scaling choices are inspired by Galilean results of `case 2b' appearing in \cite{Barducci:2019jhj}.
The contraction is carried out by taking the  $\omega\to\infty$ limit, and we obtain the extended string Carroll algebra \eqref{eq:chiral-algebra1}
supplemented with the following additional brackets
\begin{align}\label{eq:chiral-algebra2}
    \left[ M_{AB}, Z_{C} \right] &= 2\eta_{C[A}Z_{B]}\,,\hspace{1.75cm} \left[ Z_{AB'}, P_{C'} \right] = -\delta_{C'B'}Z_{A}\,,  \\
  \left[ M_{AB}, Z_{CD} \right] &= 4\eta_{[C[A}Z_{B]D]}\,, \hspace{1.55cm} \comm{Z_{AB}}{P_C}= 2\eta_{C[A}Z_{B]}\,,  \\
 \left[ M_{AB}, Z_{CD'} \right] &= 2 \eta_{C[A} Z_{B]D'} \,,\hspace{1.3cm} 
\comm{C_{AB'}}{Z_{CD}} = -2\eta_{A[C}Z_{D]B'}\,, \\
\left[ M_{A'B'}, Z_{CD'} \right] &= 2 Z_{C[A'}\delta_{B'] D'}\,, \hspace{1cm} \comm{M_{AB}}{Z_{CD'}} = 2\eta_{A[C} Z_{B] D'}\,.
\end{align}
Therefore, the resulting algebra can be written as $\mathcal{A} \oplus \mathcal{I}$ such that
\begin{align}
    \left[ \mathcal{A} ,\,  \mathcal{I} \right] \, \subseteq \, \mathcal{I}\,, \hspace{1cm} \left[ \mathcal{A}, \,\mathcal{A} \right] \, \subseteq \, \mathcal{A}\,, \hspace{1cm} \left[ \mathcal{I}, \, \mathcal{I} \right] \, \subseteq \, \mathcal{I}\,.
\end{align}
where the subalgebra $\mathcal{A}$ and ideal $\mathcal{I}$ are
\begin{align}\label{eq:ecsa}
    \mathcal{A} = \left\{ M_{AB}, M_{A'B'}, C_{AB'}, P_{A}, P_{A'}, Z_{A'}, Z_{A'B'} \right\}\,, \hspace{0.5cm} \text{and} \hspace{0.5cm} \mathcal{I} = \left\{ Z_{AB'}, Z_{AB}, Z_{A} \right\}\,.
\end{align}
Taking the quotient of $\mathcal{A}/\mathcal{I}$ we recover the extended string Carroll algebra \eqref{eq:chiral-algebra1}.

\section{Discussion}\label{sec:discussions}
We have presented two inequivalent ways of taking a Carroll limit of the flat target space of the Polyakov action, yielding two Carroll string models that we refer to as magnetic and chiral. 
Both models make use of auxiliary fields that render the action finite in the Carroll limit. In the chiral model a second set of  
auxiliary fields projects out one worldsheet chirality, leaving only left-moving string excitations. 

The magnetic model, described in Section \ref{sec:mag-carroll}, has a global symmetry algebra on the target space that coincides with the string Carroll algebra found for the CGP Carroll string model of Cardona {\it et al.} in \cite{Cardona:2016ytk}. The magnetic model and the CGP model also share the feature that the Carroll strings can carry arbitrary transverse momentum but do not move in the transverse directions, reminiscent of Carroll particles.

The dynamics of the chiral model, described in Section \ref{sec:chiralcarroll}, is less constrained and allows arbitrary left-moving transverse string excitations, which does not have a counterpart when studying Carroll particles.
The global symmetry of the chiral Carroll model is an extension of the string Carroll algebra, enlarged by a vector generator $Z_{A}$ and an anti-symmetric generator $Z_{AB}$. We show how this extended string Carroll algebra can be obtained from a contraction of the coadjoint Poincaré algebra. 

We further show how the global symmetries of chiral Carroll string model are embedded in an infinite dimensional 
symmetry algebra, which contains the infinite dimensional symmetry of the magnetic Carroll string as a subalgebra.
An analogous infinite dimensional symmetry structure was found previously for strings with Galilean symmetry in \cite{Bergshoeff:2019pij,Batlle:2016iel}. 
 
Here we have only considered closed strings and we required at least one of the transverse directions to be compactified in order to have non-trivial (left-moving) transverse string excitations in the chiral Carroll model. As a result, open strings will not have transverse modes in the models and it remains an open question whether open Carroll strings can have interesting dynamics via some other construction. 

We considered a flat target space throughout. The authors of \cite{Hartong:2021ekg,Hartong:2022dsx} considered a series expansion for a curved target space in the Galilean limit and were able to recover, amongst others, the Gomis-Ooguri model. It would be curious to see if our Carroll string models can be obtained in an analogous Carroll expansion and explore how the models couple to curved spacetime. 
Finally, relaxing the relativistic symmetry on the worldsheet might lead to interesting interplay between the target space and the worldsheet, yielding novel string spectra. 

\section*{Acknowledgements}
We thank Oscar Fuentealba, Joaquim Gomis, Troels Harmark, Jelle Hartong, Emil Have, Niels Obers, Gerben Oling, and Ziqi Yan for discussions and insightful comments.  Research supported by the Icelandic Research Fund Grant 228952-053 and by the University of Iceland Research Fund. LT would like to thank the Isaac Newton Institute for Mathematical Sciences, Cambridge, for support and hospitality during the program ``Black Holes: Bridges Between Number Theory and Holographic Quantum Information'' where work on this paper was undertaken. This work was supported by EPSRC grant no EP/K032208/1. WS acknowledges support of the Simons Foundation through the INI-Simons Postdoctoral Fellowship.

\bibliographystyle{JHEP}
\bibliography{refds}

\providecommand{\href}[2]{#2}\begingroup\raggedright\begin{thebibliography}{10}

\bibitem{Gomis:2000bd}
J.~Gomis and H.~Ooguri, {\it {Nonrelativistic closed string theory}},  {\em J.
  Math. Phys.} {\bf 42} (2001) 3127--3151
  [\href{http://arXiv.org/abs/hep-th/0009181}{{\tt hep-th/0009181}}].

\bibitem{Danielsson:2000gi}
U.~H. Danielsson, A.~Guijosa and M.~Kruczenski, {\it {IIA/B, wound and
  wrapped}},  {\em JHEP} {\bf 10} (2000) 020
  [\href{http://arXiv.org/abs/hep-th/0009182}{{\tt hep-th/0009182}}].

\bibitem{Danielsson:2000mu}
U.~H. Danielsson, A.~Guijosa and M.~Kruczenski, {\it {Newtonian Gravitons and
  D-brane Collective Coordinates in Wound String Theory}},  {\em JHEP} {\bf 03}
  (2001) 041 [\href{http://arXiv.org/abs/hep-th/0012183}{{\tt
  hep-th/0012183}}].

\bibitem{Oling:2022fft}
G.~Oling and Z.~Yan, {\it {Aspects of Nonrelativistic Strings}},  {\em Front.
  in Phys.} {\bf 10} (2022) 832271 [\href{http://arXiv.org/abs/2202.12698}{{\tt
  2202.12698}}].

\bibitem{Levy1965}
J.-M. Levy-Leblond, {\it Une nouvelle limite non-relativiste du groupe de
  poincar\'e},  {\em Annales de l'institut Henri Poincar\'e (A) Physique
  th\'eorique} {\bf 3} (1965), no.~1 1--12.

\bibitem{Bacry:1968zf}
H.~Bacry and J.~Levy-Leblond, {\it {Possible kinematics}},  {\em J.Math.Phys.}
  {\bf 9} (1968) 1605--1614.

\bibitem{Leblonde:1965}
J.-M. L{\'e}vy-Leblond, {\it Une nouvelle limite non-relativiste du groupe de
  poincar{\'e}},  {\em Annales De L Institut Henri Poincare-physique Theorique}
  {\bf 3} (1965) 1--12.

\bibitem{Gupta1966OnAA}
N.~Gupta, {\it On an analogue of the galilei group},  {\em Nuovo Cimento Della
  Societa Italiana Di Fisica A-nuclei Particles and Fields} {\bf 44} (1966)
  512--517.

\bibitem{deBoer:2017ing}
J.~de~Boer, J.~Hartong, N.~A. Obers, W.~Sybesma and S.~Vandoren, {\it {Perfect
  Fluids}},  {\em SciPost Phys.} {\bf 5} (2018), no.~1 003
  [\href{http://arXiv.org/abs/1710.04708}{{\tt 1710.04708}}].

\bibitem{deBoer:2021jej}
J.~de~Boer, J.~Hartong, N.~A. Obers, W.~Sybesma and S.~Vandoren, {\it {Carroll
  Symmetry, Dark Energy and Inflation}},  {\em Front. in Phys.} {\bf 10} (2022)
  810405 [\href{http://arXiv.org/abs/2110.02319}{{\tt 2110.02319}}].

\bibitem{Bergshoeff:2014jla}
E.~Bergshoeff, J.~Gomis and G.~Longhi, {\it {Dynamics of Carroll Particles}},
  {\em Class.Quant.Grav.} {\bf 31} (2014), no.~20 205009
  [\href{http://arXiv.org/abs/1405.2264}{{\tt 1405.2264}}].

\bibitem{Casalbuoni:2023bbh}
R.~Casalbuoni, D.~Dominici and J.~Gomis, {\it {Two interacting conformal
  Carroll particles}},  \href{http://arXiv.org/abs/2306.02614}{{\tt
  2306.02614}}.

\bibitem{deboer:2023fnj}
J.~de~Boer, J.~Hartong, N.~A. Obers, W.~Sybesma and S.~Vandoren, {\it {Carroll
  stories}},  \href{http://arXiv.org/abs/2307.06827}{{\tt 2307.06827}}.

\bibitem{Isberg:1993av}
J.~Isberg, U.~Lindstrom, B.~Sundborg and G.~Theodoridis, {\it {Classical and
  quantized tensionless strings}},  {\em Nucl. Phys. B} {\bf 411} (1994)
  122--156 [\href{http://arXiv.org/abs/hep-th/9307108}{{\tt hep-th/9307108}}].

\bibitem{Bagchi:2015nca}
A.~Bagchi, S.~Chakrabortty and P.~Parekh, {\it {Tensionless Strings from
  Worldsheet Symmetries}},  {\em JHEP} {\bf 01} (2016) 158
  [\href{http://arXiv.org/abs/1507.04361}{{\tt 1507.04361}}].

\bibitem{Bidussi:2023rfs}
L.~Bidussi, T.~Harmark, J.~Hartong, N.~A. Obers and G.~Oling, {\it
  {Longitudinal Galilean and Carrollian limits of non-relativistic strings}},
  \href{http://arXiv.org/abs/2309.14467}{{\tt 2309.14467}}.

\bibitem{Gomis:2023eav}
J.~Gomis and Z.~Yan, {\it {Worldsheet Formalism for Decoupling Limits in String
  Theory}},  \href{http://arXiv.org/abs/2311.10565}{{\tt 2311.10565}}.

\bibitem{Cardona:2016ytk}
B.~Cardona, J.~Gomis and J.~M. Pons, {\it {Dynamics of Carroll Strings}},  {\em
  JHEP} {\bf 07} (2016) 050 [\href{http://arXiv.org/abs/1605.05483}{{\tt
  1605.05483}}].

\bibitem{Bagchi:2023cfp}
A.~Bagchi, A.~Banerjee, J.~Hartong, E.~Have, K.~S. Kolekar and M.~Mandlik, {\it
  {Strings near black holes are Carrollian}},
  \href{http://arXiv.org/abs/2312.14240}{{\tt 2312.14240}}.

\bibitem{Duval:2014uoa}
C.~Duval, G.~Gibbons, P.~Horvathy and P.~Zhang, {\it {Carroll versus Newton and
  Galilei: two dual non-Einsteinian concepts of time}},  {\em
  Class.Quant.Grav.} {\bf 31} (2014) 085016
  [\href{http://arXiv.org/abs/1402.0657}{{\tt 1402.0657}}].

\bibitem{Henneaux:2021yzg}
M.~Henneaux and P.~Salgado-Rebolledo, {\it {Carroll contractions of
  Lorentz-invariant theories}},  \href{http://arXiv.org/abs/2109.06708}{{\tt
  2109.06708}}.

\bibitem{Bergshoeff:2022qkx}
E.~A. Bergshoeff, J.~Gomis and A.~Kleinschmidt, {\it {Non-Lorentzian theories
  with and without constraints}},  {\em JHEP} {\bf 01} (2023) 167
  [\href{http://arXiv.org/abs/2210.14848}{{\tt 2210.14848}}].

\bibitem{Bergshoeff:2019pij}
E.~A. Bergshoeff, J.~Gomis, J.~Rosseel, C.~\c{S}im\c{s}ek and Z.~Yan, {\it
  {String Theory and String Newton-Cartan Geometry}},  {\em J. Phys. A} {\bf
  53} (2020), no.~1 014001 [\href{http://arXiv.org/abs/1907.10668}{{\tt
  1907.10668}}].

\bibitem{Batlle:2016iel}
C.~Batlle, J.~Gomis and D.~Not, {\it {Extended Galilean symmetries of
  non-relativistic strings}},  {\em JHEP} {\bf 02} (2017) 049
  [\href{http://arXiv.org/abs/1611.00026}{{\tt 1611.00026}}].

\bibitem{Basu:2018dub}
R.~Basu and U.~N. Chowdhury, {\it {Dynamical structure of Carrollian
  Electrodynamics}},  {\em JHEP} {\bf 04} (2018) 111
  [\href{http://arXiv.org/abs/1802.09366}{{\tt 1802.09366}}].

\bibitem{Bagchi:2019xfx}
A.~Bagchi, A.~Mehra and P.~Nandi, {\it {Field Theories with Conformal
  Carrollian Symmetry}},  {\em JHEP} {\bf 05} (2019) 108
  [\href{http://arXiv.org/abs/1901.10147}{{\tt 1901.10147}}].

\bibitem{Barducci:2019jhj}
A.~Barducci, R.~Casalbuoni and J.~Gomis, {\it {Nonrelativistic $k$-contractions
  of the coadjoint Poincar\'e algebra}},  {\em Int. J. Mod. Phys. A} {\bf 35}
  (2020), no.~04 2050009 [\href{http://arXiv.org/abs/1910.11682}{{\tt
  1910.11682}}].

\bibitem{Hartong:2021ekg}
J.~Hartong and E.~Have, {\it {Nonrelativistic Expansion of Closed Bosonic
  Strings}},  {\em Phys. Rev. Lett.} {\bf 128} (2022), no.~2 021602
  [\href{http://arXiv.org/abs/2107.00023}{{\tt 2107.00023}}].

\bibitem{Hartong:2022dsx}
J.~Hartong and E.~Have, {\it {Nonrelativistic approximations of closed bosonic
  string theory}},  {\em JHEP} {\bf 02} (2023) 153
  [\href{http://arXiv.org/abs/2211.01795}{{\tt 2211.01795}}].

\end{thebibliography}\endgroup
\end{document}